\documentclass[preprint,12pt]{elsarticle}
\usepackage[numbers]{natbib}
\biboptions{numbers,sort&compress}
\usepackage{tabularx}
\usepackage[margin = 2cm]{geometry}
\usepackage{float}

\usepackage{subfigure}
\usepackage{amssymb, amsmath}

\begin{document}

\begin{frontmatter}



\title{Optimal Droop Control Strategy for Coordinated Voltage Regulation and Power Sharing in Hybrid AC-MTDC Systems}


\author[label1]{Hongjin Du}
\author[label1]{Tuanku Badzlin Hashfi}
\author[label1]{Rashmi Prasad}
\author[label1]{Pedro P. Vergara}
\author[label1]{Peter Palensky}
\author[label1]{Aleksandra Lekić}
\affiliation[label1]{organization={Delft University of Technology, Faculty of Electrical Engineering, Mathematics and Computer Science},
            addressline={Mekelweg 4},
            postcode={2628 CD}, 
            city={Delft},
            country={Netherlands}}

\begin{highlights}
\item Hierarchical optimal power flow-based droop control strategy for hybrid power systems

\item Lower generation cost while improving power-sharing performance

\item Strategy validated through time-domain simulations modified Nordic 32 test system

\item Enhanced voltage and frequency stability under dynamic conditions
\end{highlights}

\begin{abstract}
With the growing integration of modular multilevel converters (MMCs) in Multi-Terminal Direct Current (MTDC) transmission systems, there is an increasing need for control strategies that ensure both economic efficiency and robust dynamic performance. This paper presents an enhanced Optimal Power Flow (OPF) framework for hybrid AC-MTDC systems, integrating a novel droop control strategy that coordinates DC voltage and AC frequency regulation. By embedding frequency control loops into the MMCs, the method enables system-wide coordination, enhancing power sharing and improving system resilience under disturbances. The proposed strategy dynamically adjusts converter operating points to minimize generation costs and DC voltage deviations, thus balancing economic objectives with system stability. A modified Nordic test system integrated with a four-terminal MTDC grid is used to validate the approach. Optimization is performed using Julia, while the system's dynamic performance is evaluated through electromagnetic transient simulations with the EMTP software. Case studies across multiple scenarios demonstrate that the proposed method consistently achieves lower generation costs than active power control and adaptive droop control strategy while maintaining stable control characteristics. The results highlight the method’s capability to deliver cost-effective operation without compromising performance, offering a promising solution for the coordinated control of future hybrid AC-DC transmission networks.
\end{abstract}


\end{frontmatter}


\section{Introduction}
Over the past two decades, renewable energy technologies, particularly solar PV and wind power, have made remarkable progress. In 2023, global additions to renewable power capacity increased an estimated 36\% to reach 473 GW, a new record for the 22nd consecutive year. Despite these advancements, integrating high shares of renewable energy sources (RES) into utility-scale power systems remains a significant challenge. As the global share of RES exceeds 13\%, the inherent variability and unpredictability of these sources continue to strain traditional grid infrastructures, making power networks a major bottleneck in the transition toward cleaner energy \cite{REN21_2024}. In response, High Voltage Direct Current (HVDC) technology has emerged as a transformative solution, offering efficient long-distance energy transmission and facilitating the reliable integration of RES into modern power systems. The successful implementation of HVDC projects worldwide highlights its practicality and potential, as discussed in \cite{Kim2009HVDCTP}. To facilitate the coordinated integration of multiple energy sources, MTDC systems have been proposed to evolve conventional point-to-point HVDC systems. Simultaneously, MMCs have become the preferred Voltage Source Converters (VSCs), known for their modular design, control flexibility, and capacity to manage active and reactive power independently. These characteristics enable seamless interconnection with weak AC grids or even passive networks \cite{867439}. In this case, maintaining stable DC voltage is essential across all MTDC applications, as it serves as a key indicator of real-time power balance and is critical to ensuring the secure and coordinated operation of the system.

To ensure effective DC voltage regulation, existing control strategies are generally classified into two main categories: master-slave control and voltage droop control. Master-slave control achieves high precision in power-sharing but relies heavily on fast communication for transmitting active power set-points, rendering it vulnerable to single-point failures and limiting its scalability. On the other hand, voltage droop control operates on the local voltage and current measurements, allowing decentralized control without extensive communication infrastructure, which is introduced in \cite{Beerten2011VSCMS,stojkovic2020adaptive, VRANA2013137,6588621,7484690,8616804}. 

In steady-state operation, adopting localized control strategies based on power flow analysis provides a practical means of identifying cost-effective and secure operating conditions. Droop coefficients are key parameters affecting the power distribution and control performance in MTDC systems. In \cite{Beerten2011VSCMS}, a distributed DC voltage control approach is introduced, where multiple converters can collaboratively regulate the DC system voltage. After a fault event, the voltage droop-controlled converters can adjust to new operating points influencing the entire system. In \cite{stojkovic2020adaptive}, the influence of converter droop settings and DC grid topology on power sharing is analyzed, and an analytical tool is proposed to evaluate how droop control parameters affect voltage deviations during both steady-state operation and post-contingency conditions. Recent studies also have implemented distributed optimization approaches, such as Generalized Benders Decomposition (GBD) and the Alternating Direction Method of Multipliers (ADMM) into OPF framework \cite{LI2025110365}.

However, the traditional control strategy working with fixed droop gains can lead to uneven power sharing, resulting in converter overload or under-utilization, which undermines the scalability and economic efficiency of MTDC systems \cite{wang2020dc}. To address these limitations, researchers have introduced adaptive and nonlinear droop control methods, which enhance power-sharing accuracy and voltage regulation while accounting for practical factors such as sensor inaccuracies and cable resistances \cite{7967860, 9594778, Cao2013MinimizationOT}. For example, in \cite{9594778}, an adaptive reference power based droop control strategy is proposed, which improves DC voltage deviation and power-sharing under large disturbances by locally adjusting reference power through a tunable control factor.  

On the other hand, with the replacement of local synchronous generators (SGs), synchronous systems increasingly struggle with reduced rotational inertia and diminished frequency regulation capabilities. These changes can lead to a steeper rate of change of frequency (RoCoF), a deeper frequency nadir, and prolonged frequency recovery times, all of which contribute to greater instability within the system \cite{8967191}. Thus, frequency regulation within interconnected AC-DC grids is imperative for maintaining the system frequency stability. Significant efforts in developing power flow formulations for VSC-MTDC systems are commonly based on the assumption of a constant frequency. The studies conduct the droop control strategies merely based on DC variables in order to adaptively adjust the voltage in DC grids. However, these approaches neglect the inherent coupling between DC variables and AC frequency, as well as the critical role of power injections at the point of common coupling (PCC) in shaping overall system behavior. Furthermore, they often assume uniform AC frequency across the entire AC network, which oversimplifies real-world operating conditions. The control performance in frequency regulation has been analyzed in previous studies, primarily based on simplified network configurations and microgrids \cite{8103883, XIAO2021106604, lekic2020initialisation}. In \cite{8103883}, a branch-based load flow approach for hybrid AC/DC microgrids, in which the DC subgrid and its effect on the inter-iteration frequency updating. A coordinated frequency regulation scheme that enables spinning reserve sharing is proposed in \cite{XIAO2021106604}, while \cite{lekic2020initialisation} utilizes the power flow solution as an initialization method for hybrid AC-DC power systems, providing a foundation for subsequent electromagnetic or harmonic stability analyses. However, existing models do not accurately reflect the topology of realistic AC-MTDC interconnections or fully exploit the decoupled control capabilities of VSCs.

Motivated by the aforementioned problems, a novel and comprehensive approach is developed to enhance the OPF-based droop control strategy considering both voltage and frequency in AC-MTDC systems. The main contributions of this paper are as follows: (1) A coordinated droop control strategy is proposed, enabling simultaneous regulation of DC voltage and AC frequency by embedding frequency control loops into MMCs, typically implemented through secondary control mechanisms with PI controllers; (2)  An economic operation framework is developed, which dynamically adjusts system operating points to minimize generation costs and DC voltage deviations while ensuring stable MTDC grid performance; (3) The proposed strategy is validated through simulations on the Nordic 32 system integrated via MMCs with a four-terminal DC grid, demonstrating its effectiveness and feasibility under various operating scenarios.

The remainder of this paper is organized as follows. The configuration of MMC stations and the conventional voltage droop control scheme are reviewed in Section \uppercase\expandafter{\romannumeral2}. The proposed OPF-based droop control strategy is presented in Section \uppercase\expandafter{\romannumeral3}. The feasibility of the proposed control scheme and its impact on the stability of the test system are verified through simulations in Section \uppercase\expandafter{\romannumeral4}, and Section \uppercase\expandafter{\romannumeral5} draws conclusions.

\section{Configuration of MMCs and Conventional Voltage Droop Control}

\subsection{Overview of Configuration of MMCs}

\begin{figure}[htbp]
\centerline{\includegraphics[scale=0.35]{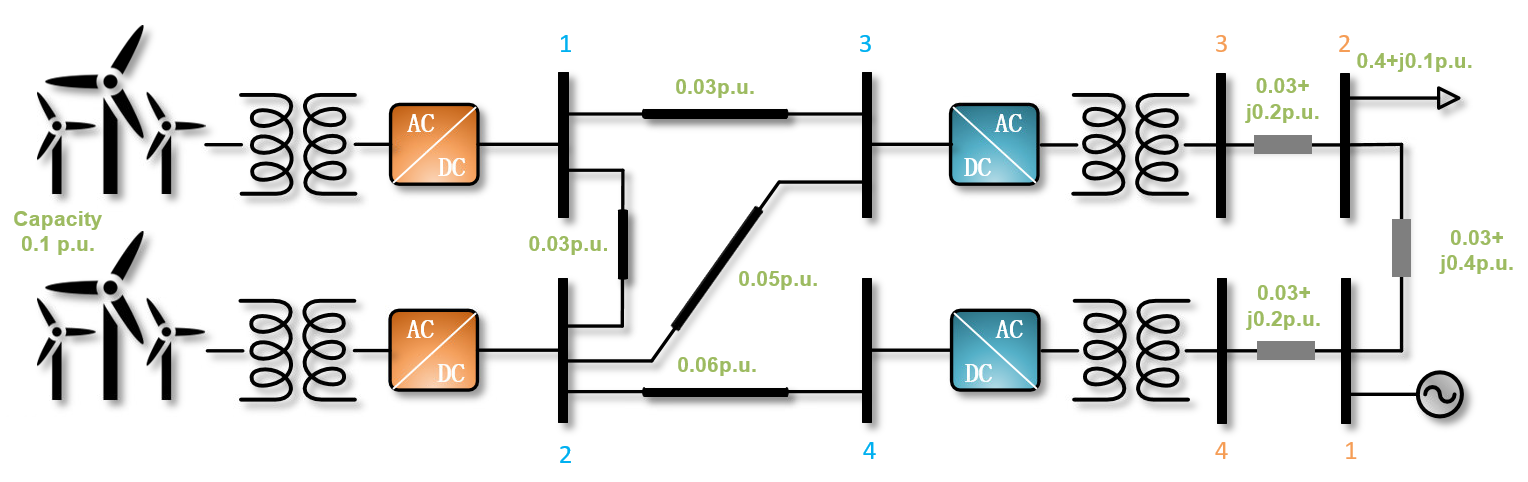}}
\caption{Configuration of a 4-terminal AC-MTDC system.}
\label{topology}
\end{figure}

As the most widely adopted type of VSC, MMCs play a critical role in large-scale MTDC systems by enabling efficient bidirectional conversion between AC and DC power. Figure~\ref{topology}  shows a typical configuration for a four-terminal AC-MTDC transmission system connected to offshore wind farms. 

\begin{figure}[htbp]
\centerline{\includegraphics[scale=1.2]{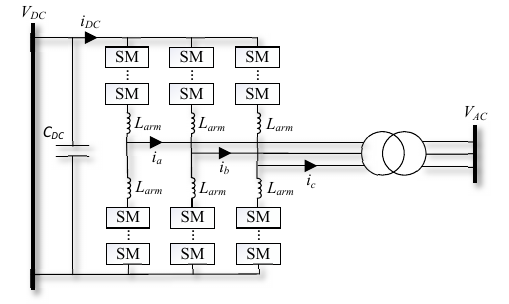}}
\caption{Overview of MMC design.}
\label{mmc}
\end{figure}

The fundamental structure of an MMC comprises multiple half-bridge submodules (SMs) connected in series to form arms and then further combined in legs, which create a converter, as shown in Figure~\ref{mmc}. Unlike conventional two-level or three-level VSCs, which rely solely on pulse-width modulation (PWM) for voltage synthesis, an MMC primarily synthesizes its output voltage by inserting or bypassing SMs in each arm, forming a stepped waveform. PWM techniques may still be applied within individual SMs to enhance waveform quality or control performance. The number of active SMs ($N_{active}$) at any time determines the generated voltage:
\begin{equation} 
U_{\text{arm}} = \sum_{i=1}^{N} S_i U_{\text{SM},i}, \label{eq:MMC_voltage} 
\end{equation}
where $S_i\in\{0,1\}$ is the switching state of the $i$-th SM, and $U_{\text{SM},i}$ is its capacitor voltage. The modular index M represents the fraction of SMs engaged in voltage synthesis:
\begin{equation} 
M = \frac{N_{\text{active}}}{N_{\text{total}}}. \label{eq:MMC_modular_index} 
\end{equation}

By dynamically adjusting the number of inserted submodules $N_{\text{active}}$, MMCs can approximate a sinusoidal waveform more closely than conventional VSCs. This results in smoother output voltages, reduced switching losses, and improved voltage scalability. Additionally, the inherent characteristics of MMCs contribute to harmonic mitigation without additional filtering, thereby potentially reducing overall system complexity. Furthermore, optimizing the modulation index reduces component stress and energy losses and contributes to overall cost savings. In this paper, the modular settings of MMCs are derived from the optimization results and subsequently applied in time-domain simulations, ensuring consistency between planning and dynamic performance evaluation.

\subsection{Conventional Voltage Droop Control}

\begin{figure}[htbp]
\centerline{\includegraphics[scale=0.7]{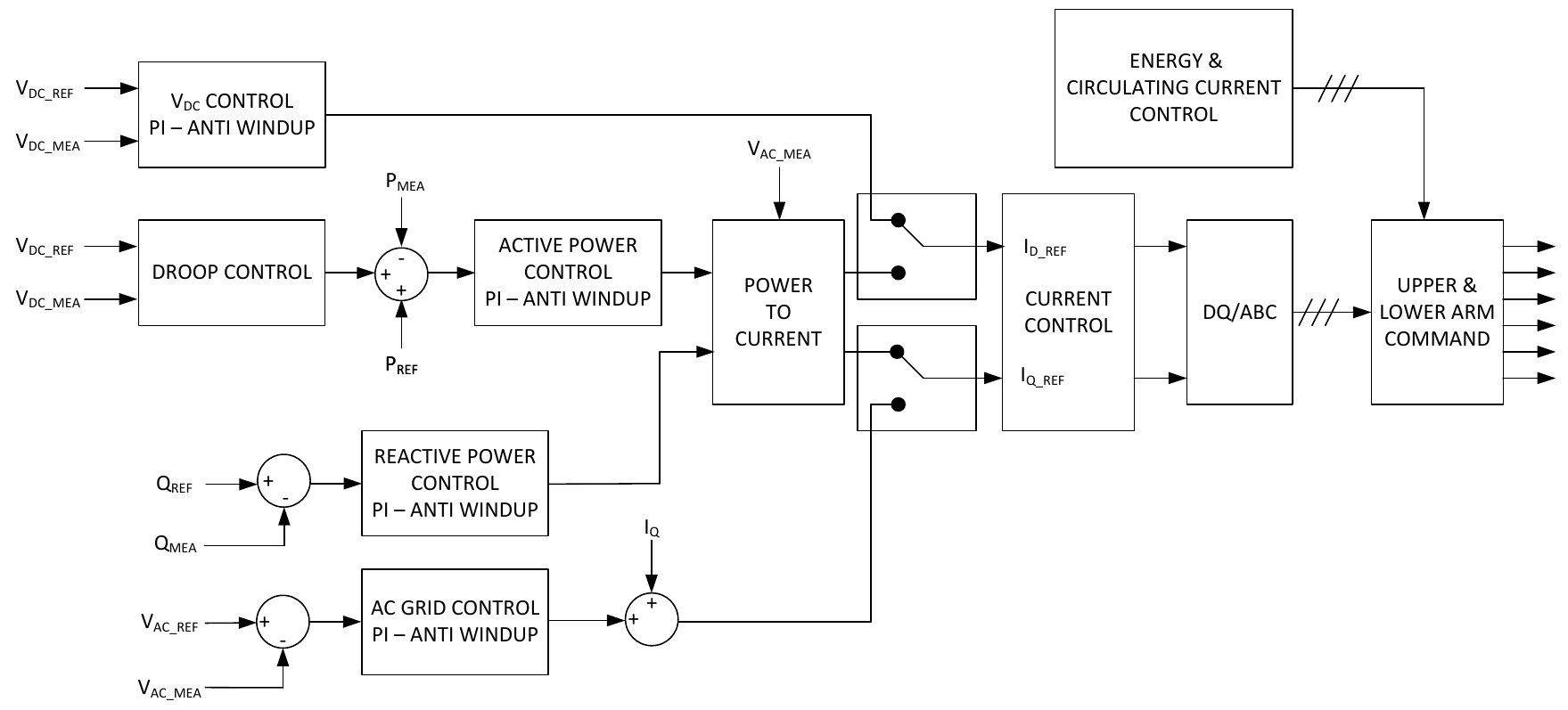}}
\caption{MMC operation diagram.}
\label{operation modes}
\end{figure}

A typical MMC employs a hierarchical double-loop control structure, as illustrated in Figure~\ref{operation modes}. The outer loop, often called the secondary control, regulates the DC voltage and active power exchange based on a predefined droop characteristic. This layer enables coordinated system-wide regulation across multiple terminals, while the inner loop comprising current controllers ensures fast tracking of reference values by modulating converter output in real-time. In addition, the energy control loop monitors the energies of the submodule capacitors and adjusts the internal circulating current to maintain the desired energy levels, ensuring convergence and stability.

The droop control enables decentralized power-sharing and voltage regulation without requiring an extensive communication infrastructure. The active power injection at the DC side of the $i$-th converter can be described by the droop equation:
\begin{equation}
    P_{dc,i}-P_{dc,0}+1/k_{droop}\left(U_{dc, i}-U_{dc,0}\right)=0,
    \label{droopfunction}
\end{equation}
where $k_{droop}$ denotes the droop coefficient; $P_{dc, i}$ and $P_{dc,0}$ are the actual and reference power injected into the DC grid; and $U_{dc, i}$ and $U_{dc,0}$ are the actual and reference voltage, respectively. 

In an ideal lossless MTDC grid, the unbalanced power arising from fluctuations or disturbances is naturally distributed proportionally among all participating VSC stations to their droop gains. This characteristic offers a simple yet powerful means for voltage regulation and power-sharing. The operator can thus finely tune the droop coefficient $k_{droop}$ to control how each converter contributes to voltage support and frequency stabilization following disturbances. However, while smaller droop coefficients increase converter participation and responsiveness, overly aggressive settings can introduce instability or oscillations due to tightly coupled control loops.

The selection of $k_{droop}$ is therefore critical, as it affects steady-state power distribution and impacts the system's dynamic response and stability margins. In practical scenarios, converters operate under diverse conditions, including varying inertia, load profiles, and system topologies. These complexities necessitate a robust and flexible control design.

To address these challenges, this paper proposes a novel coordinated droop control strategy that enhances the conventional approach by simultaneously regulating DC voltage and stabilizing AC frequency. This approach dynamically adjusts droop settings in accordance with operating conditions, thereby improving post-disturbance recovery and ensuring smoother transitions to new steady states. By embedding system-level optimization into the control framework, the proposed strategy achieves effective voltage regulation, fair power-sharing, and improved dynamic stability, without relying on centralized communication or exhaustive computational resources. Further details are discussed in the following section. 

\section{Proposed Control Framework}

As mentioned before, traditional droop strategies often overlook the interaction between voltage regulation and power dispatch, particularly under varying system conditions. To overcome these limitations, we develop a coordinated control framework and an optimal droop-based control model grounded in the AC/DC power flow formulation, as originally introduced in \cite{5589968}. This integration enables converter control actions to be directly embedded within the steady-state optimization process, yielding more accurate and realistic operating points for the hybrid power system.

\subsection{Formulation of the Optimization Problem}
In this paper, the optimization problem is formulated as a two-stage process aimed at minimizing both generation cost and DC voltage deviation. The corresponding objective functions are defined as:
\begin{align}
Obj_1 = \min \sum_{i=1}^M (\alpha_iP_{Gi}^2 + \beta_iP_{Gi} + \gamma_i), \\
Obj_2 = \min \sum_{j=1}^N (U_{dc,j} - U_{rated,j})^2 ,
\end{align}
where $Obj_1$ aims to reduce the generation cost, and $Obj_2$ minimizes the voltage deviation across DC nodes. Here, $P_{Gi}$ denotes the active power generation at generator $i$ in set $M$, while $\alpha_i$, $\beta_i$, and $\gamma_i$ are the corresponding cost coefficients. $U_{dc,j}$ and $U_{rated,j}$ represent the actual and rated DC voltages at DC node $j$ in set $N$, respectively.

In conventional OPF formulations, generation cost minimization (\(Obj_1\)) is typically prioritized, focusing on steady-state economic performance. However, while it also addresses such dynamics indirectly, the unique role of MMCs in hybrid systems requires further consideration. Specifically, MMCs can participate in power sharing by adjusting their power injections in response to DC voltage variations. Motivated by this, the proposed strategy incorporates Objective 2 on minimizing DC voltage deviation (\(Obj_2\)), alongside conventional cost minimization. This integrated objective helps to regulate the DC voltage toward its nominal level under various operating conditions, thereby improving the dynamic response of the system. It should be emphasized that the proposed formulation does not adopt a conventional multi-objective optimization framework. Instead, it adopts a hierarchical structure in which ensuring DC voltage stability is given precedence over cost minimization, particularly under dynamic operating conditions. 

For the AC network, the power flow equations are expressed as:
\begin{align}
P_{i} = U_i \sum\limits_{j=1}^n U_j[G_{ij}cos(\delta_i-\delta_j)+B_{ij}sin(\delta_i-\delta_j)],\\
Q_{i} = U_i \sum\limits_{j=1}^n U_j[G_{ij}sin(\delta_i-\delta_j)-B_{ij}cos(\delta_i-\delta_j)],\\
P_{Gi} - P_{Di} - P_{i} = 0,\quad
Q_{Gi} - Q_{Di} - Q_{i} = 0,
\end{align}
where $P_i$ and $Q_i$ are the active and reactive power at AC bus $i$, $P_{Di}$ and $Q_{Di}$ are the corresponding demands, and $P_{Gi}$ and $Q_{Gi}$ represent the generated powers. $U_i$ is the bus voltage magnitude, $G_{ij}$ and $B_{ij}$ are the conductance and susceptance between AC buses $i$ and $j$, and $\delta_i$ and $\delta_j$ are the voltage phase angles.

The DC grid is modeled using the following set of equations: 
\begin{align}
I_{dc,i}=\sum\limits_{j=1, j \neq i}^n Y_{dc,ij} (U_{dc,i}-U_{dc,j}),\\
P_{dc,i} = 2U_{dc,i}I_{dc,i},
\end{align}
where $I_{dc,i}$ is the injected DC current at DC bus $i$, and $Y_{dc,ij}$ is the admittance between DC buses $i$ and $j$.

Converter station power losses, denoted as $P_{loss, i}$, include three components: no-load losses, linear losses, and quadratic losses, based on the reactor current $I_{c, i}$. The empirical coefficients $a$, $b$, and $c$ are listed in Table~\ref{powerloss}. The converter losses and current relationships are modeled as follows:
\begin{align}
P_{loss,i} = a+b*I_{c,i}+c*I_{c,i}^2, \\
P_{c,i}^2 + Q_{c,i}^2 - 3U_{c,i}I_{c,i} = 0
\end{align}

\begingroup
\begin{table}[h!] 
\centering
\caption{VSC power loss coefficients \cite{5589968}.}
\footnotesize 
\renewcommand{\arraystretch}{1.25}
\begin{tabularx}{0.45\textwidth}{ >{\raggedright\arraybackslash}X >{\centering\arraybackslash}X >{\centering\arraybackslash}X >{\centering\arraybackslash}X }
 &\textbf{a}&\textbf{b}&\textbf{c}\\ [1pt]
\hline
Rectifier & 0.011 & 0.003 & 0.004\\[1pt]
Inverter & 0.011 & 0.003 & 0.007\\[1pt]
\hline
\end{tabularx}
\label{powerloss}
\end{table}
\endgroup

Finally, a set of operational constraints is applied to guarantee the physical feasibility and secure operation of the system:
\begin{align}
     U_i^{min} \leq U_i \leq U_i^{max}, \quad 
    \delta_i^{min} \leq \delta_i \leq \delta_i^{max}, \\\quad
     P_{Gi}^{min} \leq P_{Gi} \leq P_{Gi}^{max}, \qquad
     Q_{Gi}^{min} \leq Q_{Gi} \leq Q_{Gi}^{max},\\ \quad
     P_{dc,i}^{min} \leq P_{dc,i} \leq P_{dc,i}^{max}.
\end{align}
These constraints enforce acceptable voltage magnitudes and phase angles at AC buses, along with active and reactive power limits of generators. For the DC side, converter injection limits are imposed to reflect the technical boundaries of power exchange between AC and DC grids. Together, these bounds form the foundation for enabling the OPF to yield solutions that are not only optimal but also practically implementable in real-time simulations.

\subsection{Proposed Droop Control Strategy}
\begin{figure}[ht!]
\centerline{\includegraphics[scale=1.0]{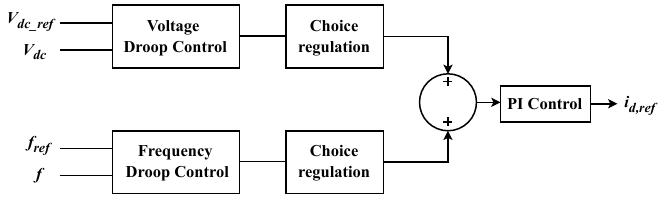}}
\caption{Voltage-frequency control in the outer control of loop of the d-q control system.}
\label{control loop}
\end{figure}

This section introduces the proposed droop control strategy, which is integrated into the optimization process described previously. Given the complexities and limitations of AC systems connected to DC grids, it is essential to account for system frequency regulation in the optimal power flow process, especially for weak AC systems. This can be achieved by incorporating a frequency regulating block into the outer control loop of the d-q control system replacing the conventional DC voltage controller, which is shown in Figure\ref{control loop}. The output from the frequency controller is added to the voltage droop result to generate the power reference for each converter. It is noted that the frequency controller does not modify the coefficients of the voltage droop characteristic. Instead, it serves as a supplementary mechanism to the voltage controller, ensuring that the frequency at each associated AC bus remains aligned with its reference. In this study, the focus shifts from the system's average frequency to the frequency at the PCC points between the AC system and the converters. This localized perspective enables a more precise analysis of AC-DC interactions, where frequency deviations directly drive power-sharing decisions through the converter’s droop control mechanism.

\begin{figure}[htbp]
\centerline{\includegraphics[scale=1.0]{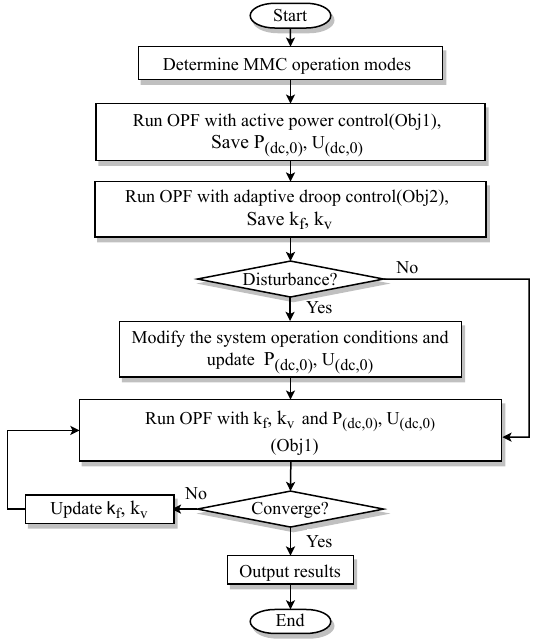}}
\caption{Flowchart of the proposed strategy.}
\label{flowchart}
\end{figure}
Building on our previous work, which presented an enhanced OPF-based voltage droop control strategy \cite{10863591}, we propose an optimized approach that integrates frequency regulation into AC-MTDC grids, aiming to improve steady-state power sharing and maintain voltage and frequency stability under both normal and disturbed operating conditions.

The flowchart of the proposed strategy is illustrated in Figure\ref{flowchart}. The process begins with the operation of MMCs in active power control mode, aimed at minimizing generation cost (Objective 1). This step determines the active power and voltage set points, \(P_{dc,0}\) and \(U_{dc,0}\) at the DC buses, which are used as reference values for the droop control. In the second stage, the control shifts to adaptive droop mode, where the frequency and voltage droop coefficients \(k_v\) and \(k_f\) are treated as variables with the goal of minimizing DC voltage deviation (Objective 2). As noted in \cite{9178970}, frequent adjustments of droop coefficients can potentially impact the stability of the AC-DC system, possibly causing undesirable interactions between the terminals. To address this issue, droop coefficients are initially set and only updated following disturbances.

After a disturbance, the system's operating conditions are updated through the OPF process. The updated droop coefficients \(k_v\) and \(k_f\), along with the revised active power and voltage set points \(P_{dc,0}\) and \(U_{dc,0}\), are then used as inputs to the OPF with Objective 1. If the optimization converges, Objective 1 is applied in the final step to update the droop characteristics. Otherwise, the droop coefficients \(k_v\) and \(k_f\) from the previous conditions are revised, and the final optimization step is repeated.

The governing equation for this process is as follows:
\begin{equation}
    P_{dc,i}-P_{dc,0}+1/k_v\left(U_{dc, i}-U_{dc,ref}\right) + 1/k_f\left(f_i-f_{i,ref}\right)=0,  
\end{equation} 
where \(f_i\) denotes the steady-state frequency and \( f_{i,\text{ref}} \) refers to the reference frequency value at PCC points. The parameters \(k_v\) and \(k_f\) represent the voltage and frequency droop coefficients, respectively, governing the converter’s active power response to deviations in DC voltage and AC frequency.

This approach ensures the simultaneous optimization of generation cost minimization and DC voltage stability during dynamic conditions. By integrating both voltage and frequency regulation into the optimization framework, the proposed strategy offers a well-balanced solution that enhances the operation of the AC-MTDC grid in both steady-state conditions and during transient disturbances.

\section{Simulation Results}
\begin{figure}[htbp]
    \vspace{-20mm} 
    \centering
    \subfigure[Topology of Nordic 32 test system]{
        \includegraphics[width=0.65\textwidth]{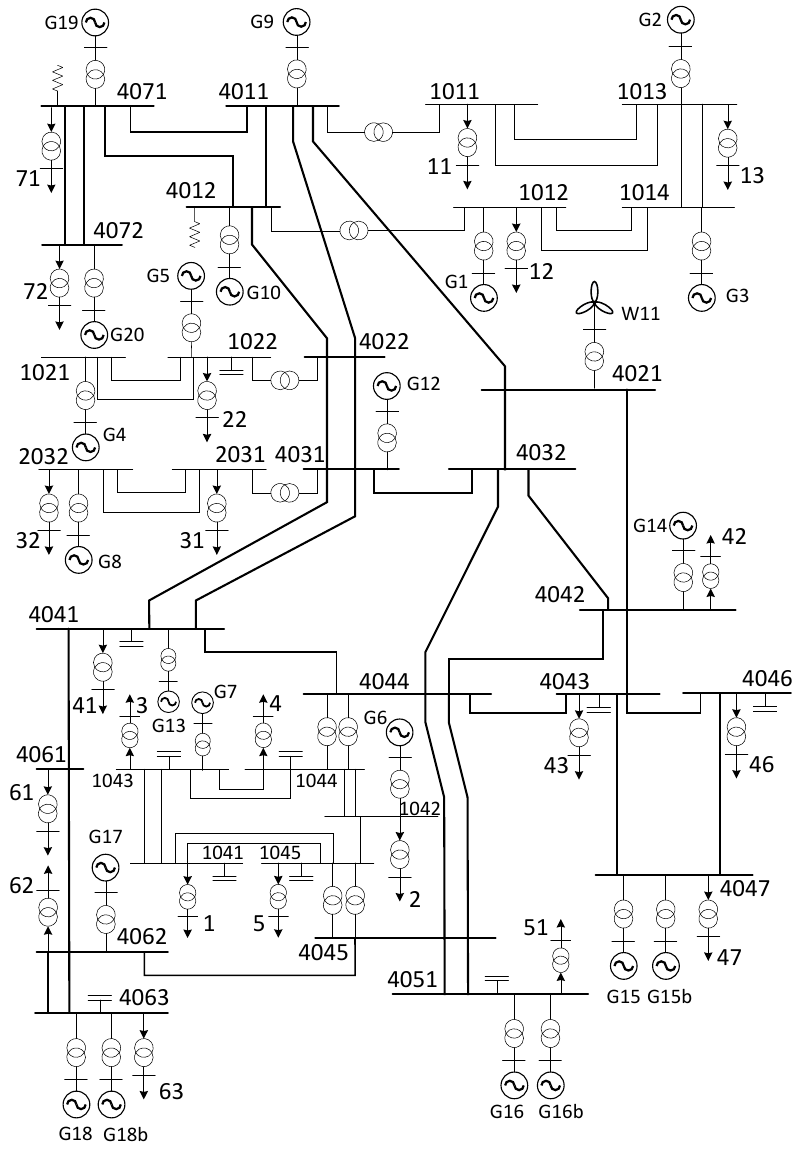}}
    \\ 
    \subfigure[Embedded MTDC system]{
        \includegraphics[width=0.4\textwidth]{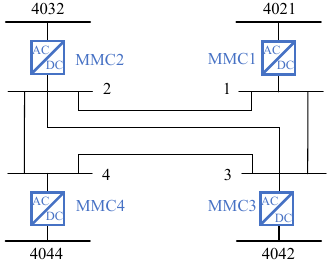}}
    \caption{Modified nordic test system with a 4-terminal MTDC system \cite{8950069, van2015test}.}
    \label{nordic}
\end{figure}

To evaluate the proposed control strategy, a modified Nordic 32 test system integrated with a 4-terminal DC grid and a wind farm, as illustrated in Figure~\ref{nordic}, is modeled in MATPOWER format for OPF simulations. The case system parameters utilized in this paper are sourced from \cite{8950069} and \cite{van2015test}, and the system works under the operating point B settings. The nominal power of the whole system is $S_{nom} = 100$ MVA while the DC nominal voltage is $V_{nom} = 200$ kV. 

Following that, the proposed control strategy is executed, building upon the power model as its cornerstone. The optimization process is implemented using a forked version of the PowerModelsACDC.jl package \cite{8636236}, which is further developed in our work \cite{hj_modifiedpkg_2025}. To facilitate comparative analysis, three dedicated branches are established within the repository: (i) active power control, (ii) adaptive droop control, and (iii) the proposed droop control strategy. Each branch independently formulates and solves the respective optimization problem, generating the operating points for the hybrid AC-MTDC system. The nonlinear optimization problem is solved using the Interior Point OPTimizer (IPOPT) solver, integrated through the JuMP modeling framework. Subsequently, the obtained results are cross-validated in EMTP to assess the system's transient performance under different control strategies.

Three representative scenarios are selected to evaluate the performance of the control strategies, each designed based on realistic operational conditions and structured with increasing levels of disturbance severity to test the robustness of the control mechanisms.
\begin{enumerate}
    \item \textbf{Scenario 1:} The system operates under normal condition.
    \item \textbf{Scenario 2:} A fault is initiated at Generator 16, leading to its subsequent disconnection from the system.
    \item \textbf{Scenario 3:} A fault is applied to MMC 4, resulting in its disconnection from the associated AC grid.
\end{enumerate}

Additionally, three control strategies are implemented and comparatively analyzed to evaluate system performance under different operating conditions, as mentioned previously. These strategies include:
\begin{enumerate}
    \item \textbf{Active power control:} All MMCs operate in constant power control mode, without adaptive response to voltage or frequency fluctuations. 
    \item \textbf{Adaptive droop control:} All MMCs are configured with voltage droop control. The droop coefficients ($k_v$) are determined by OPF calculations and are allowed to vary according to system conditions.
    \item \textbf{Proposed droop control:} MMCs 1 and 2 are equipped with both voltage and frequency droop control, while MMCs 3 and 4 implement only voltage droop control. Similarly to adaptive droop control, the droop coefficients are derived from OPF results.
\end{enumerate}

The droop coefficients obtained by solving the OPF problem in various scenarios are summarized in Table~\ref{kdroop_adaptive}, Table~\ref{kdroop}, and Table~\ref{kfdroop}, corresponding to the adaptive droop control and the proposed control strategy, respectively. It should be noted that in the proposed control mode, MMCs 1 and 2 are responsible for both voltage and frequency regulation, while MMCs 3 and 4 operate solely in voltage control mode. This configuration is designed to facilitate a comparative analysis of system performance under different control schemes. The range for the droop coefficients is carefully set between 0.001 and 1 to maintain the system stability and ensure a balanced trade-off between dynamic response and voltage regulation. 

The droop coefficients $k_v$ in both control strategies exhibit distinct adjustments in response to different scenarios. Under adaptive droop control, the values for the MMCs are fixed at their limits across all scenarios, which may not be optimal for voltage regulation under dynamic conditions, as it limits the flexibility of the system. In contrast, the proposed droop control allows for more varied values, with the coefficients generally falling within a middle range. This suggests more adaptive and flexible voltage regulation, helping to prevent the system from hitting the control limits and enhancing the ability to respond to system disturbances. On the other hand, the frequency droop coefficients $k_f$ under the proposed droop control show significant variation across different scenarios adjusting the variance of the frequency references.  In Scenario 2, where the system experiences a significant fault, the system may initially face instability due to the sudden loss of generation. By reducing $k_f$, the system enhances its responsiveness to frequency changes, compensating for the disturbance by more actively regulating the frequency to prevent further instability. In Scenario 3, the droop coefficients exhibit distinct adjustments across the remaining MMCs following the outage of MMC 4. Specifically, the droop coefficients $k_v$ and $k_f$ of MMC 1 and MMC 3 increase, indicating a reduction in their power-sharing contribution in response to voltage and frequency deviations. In contrast, MMC 2 shows a decrease in droop coefficients, thereby assuming a larger share of the load compensation. This redistribution of control effort reflects an adaptive rebalancing mechanism within the system, ensuring that the converters respond proportionally based on their capacity and network location. This differentiation helps maintain the support of voltage and frequency while avoiding excessive response from any single converter under constrained operating conditions.

The final optimization results of minimizing generation costs are presented in Table~\ref{objective}. It is observed that active power control generally exhibits superior cost-effectiveness relative to droop control, primarily owing to its capacity to directly optimize power flow without the inherent constraints associated with droop-based regulation. The implementation of droop characteristics can lead to increased generation costs, especially in systems that aim to satisfy multiple operational objectives. However, it is crucial to recognize that droop control strategies offer valuable advantages in enhancing system stability during disturbed conditions by offering reserve margins and supporting robust voltage and frequency regulation. 

Optimization results demonstrate that the proposed droop control strategy consistently achieves lower generation costs than adaptive voltage droop control, indicating better cost-effectiveness in the tested scenarios. This outcome may be attributed to the adaptive scheme's tendency to modify power references dynamically in response to disturbances, which can lead to suboptimal dispatch decisions and slightly higher generation costs under certain conditions. Alternatively, the proposed method offers a more cost-effective approach by enabling a more gradual response to operational changes. Furthermore, the objective values achieved under the proposed control strategy remain relatively stable across all tested scenarios, demonstrating its ability to consistently maintain cost efficiency. 

\begingroup
\begin{table}[!htbp]
\centering
\caption{$k_{v}$ in different scenarios under adaptive droop control.} \label{kdroop_adaptive}
\begin{minipage}{0.8\textwidth}
\footnotesize 
\renewcommand{\arraystretch}{1.5}
\begin{tabularx}{\textwidth}{ >{\raggedright\arraybackslash}X >{\centering\arraybackslash}X >{\centering\arraybackslash}X >{\centering\arraybackslash}X }
 & \textbf{Scenario 1} & \textbf{Scenario 2} & \textbf{Scenario 3}\\ [1pt]
\hline
MMC 1 & 1.0000 & 1.0000 & 1.0000\\[1pt]
MMC 2 & 1.0000 & 1.0000 & 1.0000\\[1pt]
MMC 3 & 1.0000 & 1.0000 & 0.0010\\[1pt]
MMC 4 & 0.0010 & 0.0010 & -\\[1pt]
\hline
\end{tabularx}
\end{minipage}
\end{table}
\endgroup

\begingroup
\begin{table}[!htbp]
\centering
\caption{$k_{v}$ in different scenarios under proposed droop control.} \label{kdroop}
\begin{minipage}{0.8\textwidth}
\footnotesize 
\renewcommand{\arraystretch}{1.5}
\begin{tabularx}{\textwidth}{ >{\raggedright\arraybackslash}X >{\centering\arraybackslash}X >{\centering\arraybackslash}X >{\centering\arraybackslash}X }
 & \textbf{Scenario 1} & \textbf{Scenario 2} & \textbf{Scenario 3}\\ [1pt]
\hline
MMC 1 & 0.2936 & 0.4863 & 0.9999\\[1pt]
MMC 2 & 0.1599 & 0.4896 & 0.3594\\[1pt]
MMC 3 & 0.3430 & 0.9503 & 0.9999\\[1pt]
MMC 4 & 0.2926 & 0.9560 & -\\[1pt]
\hline
\end{tabularx}
\end{minipage}
\end{table}
\endgroup

\begingroup
\begin{table}[!htbp] 
\centering
\caption{$k_{f}$ in different scenarios under proposed droop control.} \label{kfdroop}
\begin{minipage}{0.8\textwidth}
\footnotesize 
\renewcommand{\arraystretch}{1.5}
\begin{tabularx}{\textwidth}{ >{\raggedright\arraybackslash}X >{\centering\arraybackslash}X >{\centering\arraybackslash}X >{\centering\arraybackslash}X }
 & \textbf{Scenario 1} & \textbf{Scenario 2} & \textbf{Scenario 3}\\ [1pt]
\hline
MMC 1 & 0.5303 & 0.5065 & 0.9995\\[1pt]
MMC 2 & 0.5392 & 0.5052 & 0.5309\\[1pt]
\hline
\end{tabularx}
\end{minipage}
\end{table}
\endgroup

\begingroup
\begin{table}[!htbp] 
\centering
\caption{Objective values($\times 10^{6}$ MW) in different scenarios.} \label{objective}
\footnotesize 
\renewcommand{\arraystretch}{1.5}
\begin{tabularx}{\textwidth}{ >{\raggedright\arraybackslash}p{5cm}  >{\centering\arraybackslash}X >{\centering\arraybackslash}X >{\centering\arraybackslash}X }
 & \textbf{Scenario 1} & \textbf{Scenario 2} & \textbf{Scenario 3}\\ 
\hline
Active Power Control & 7.1559 & 7.4725 & 7.1251 \\[1pt]
Adaptive Voltage Droop & 7.2828 & 7.6842 & 7.2801 \\[1pt]
Proposed Voltage Droop & 7.2785 & 7.6835 & 7.2359\\[1pt]
\hline
\end{tabularx}
\end{table}
\endgroup

\begin{figure}[H]
    \centering
    \subfigure[DC voltages at MMC 1]{
        \includegraphics[width=0.48\textwidth]{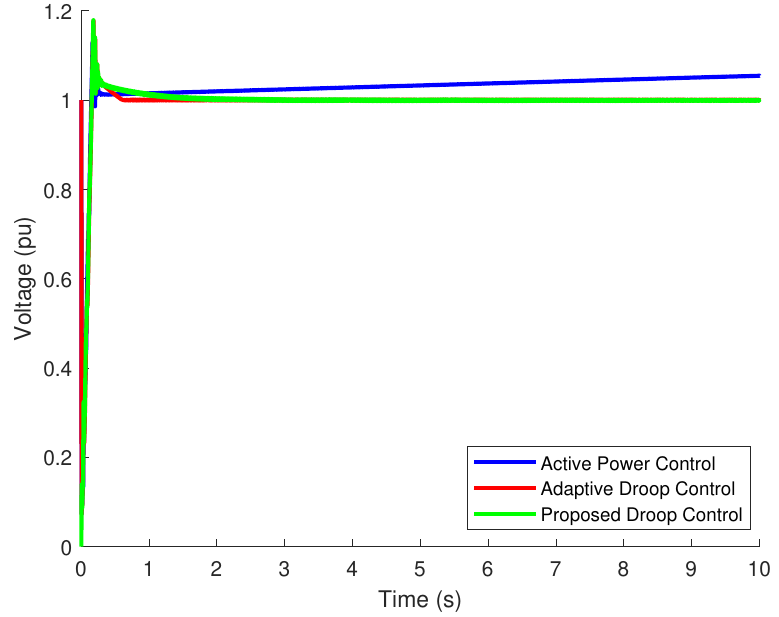}}%
    \hfill
    \subfigure[DC voltages at MMC 2]{
        \includegraphics[width=0.48\textwidth]{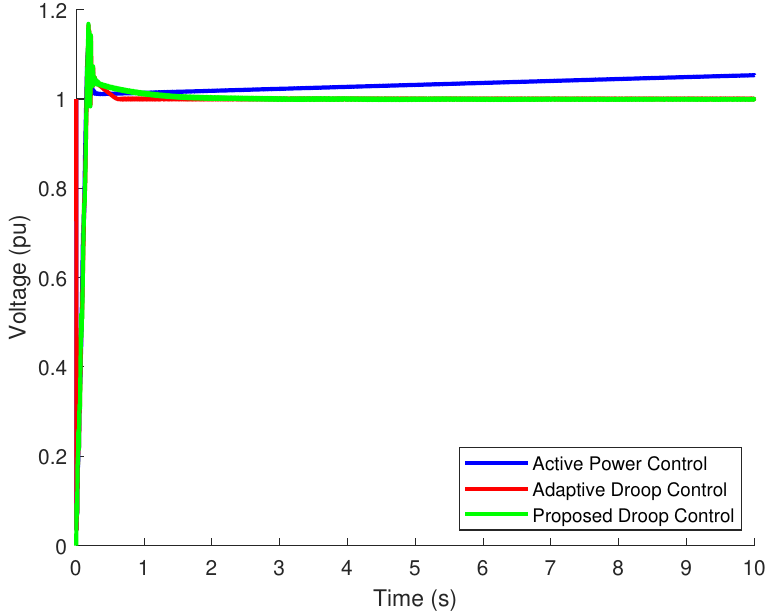}}
     \subfigure[DC voltages at MMC 3]{
        \includegraphics[width=0.48\textwidth]{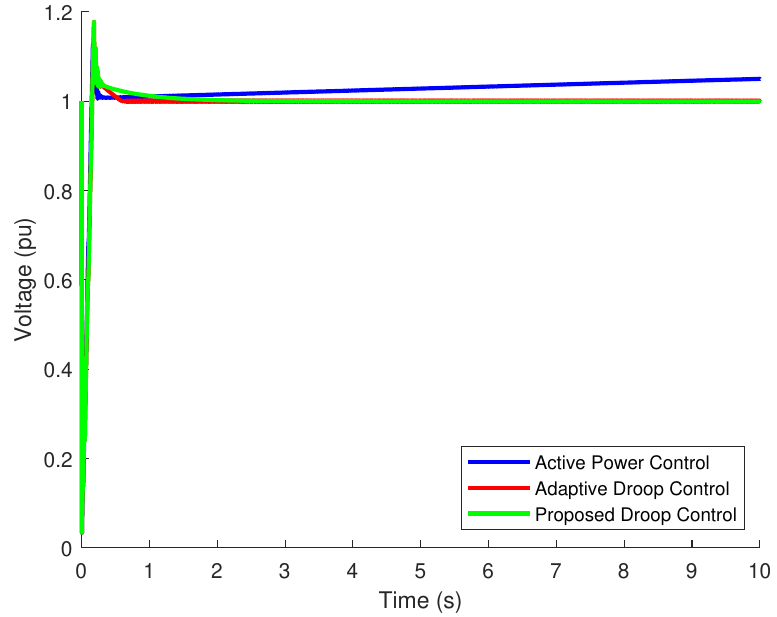}}
    \subfigure[DC voltages at MMC 4]{
        \includegraphics[width=0.48\textwidth]{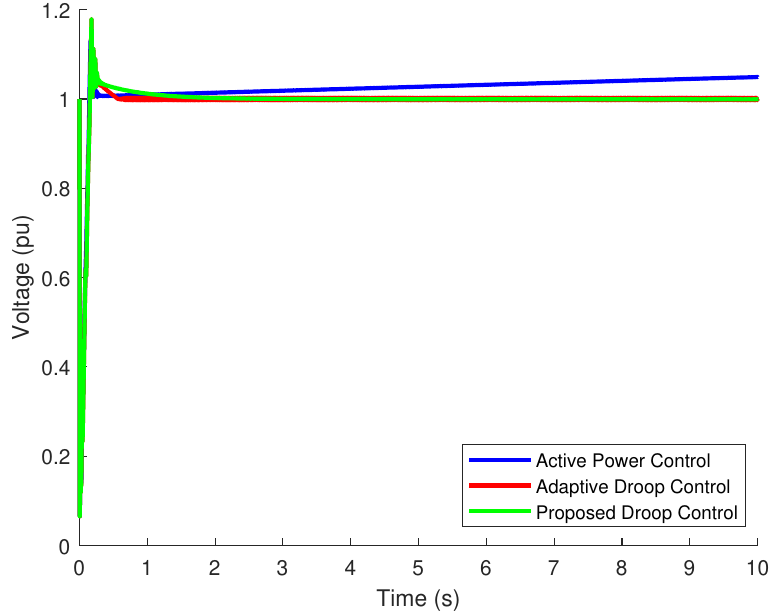}}
    \caption{DC voltages in Scenario 1 under various control strategies.}
    \label{voltage_scenario1}
\end{figure}

\begin{figure}[!htbp]
    \centering
    \subfigure[AC frequencies at MMC 1]{
        \includegraphics[width=0.48\textwidth]{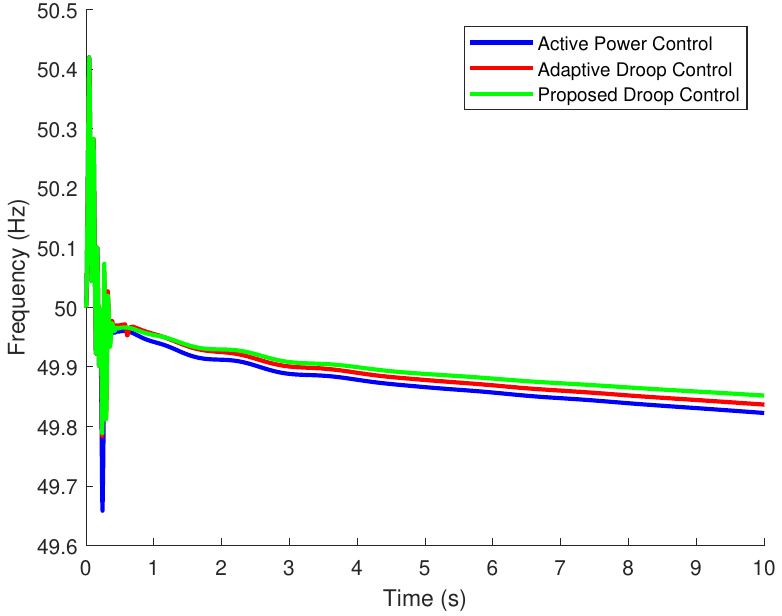}}
    \subfigure[AC frequencies at MMC 2]{
        \includegraphics[width=0.48\textwidth]{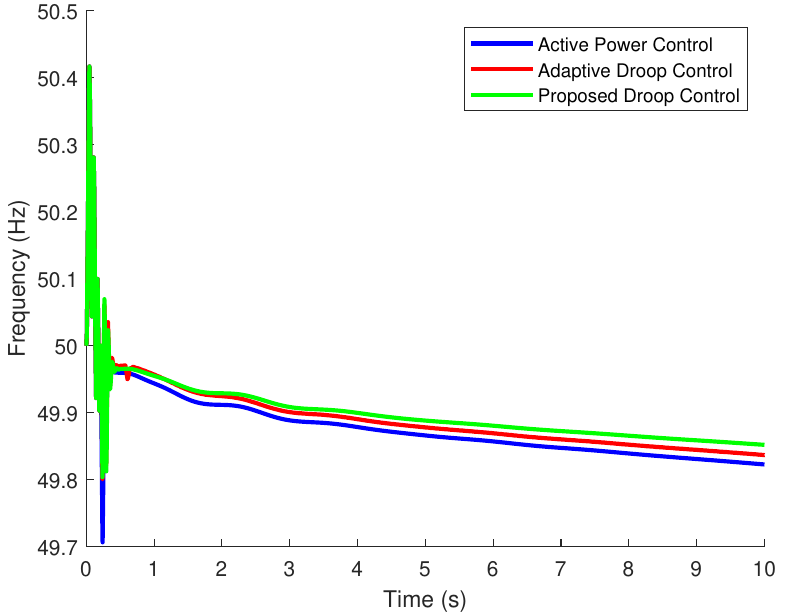}}
    \subfigure[AC frequencies at MMC 3]{
        \includegraphics[width=0.48\textwidth]{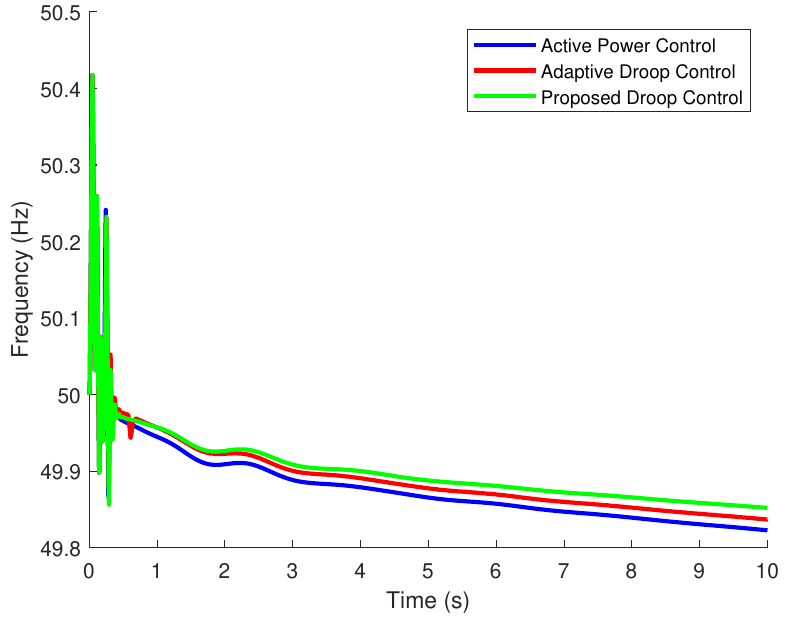}}
    \subfigure[AC frequencies at MMC 4]{
        \includegraphics[width=0.48\textwidth]{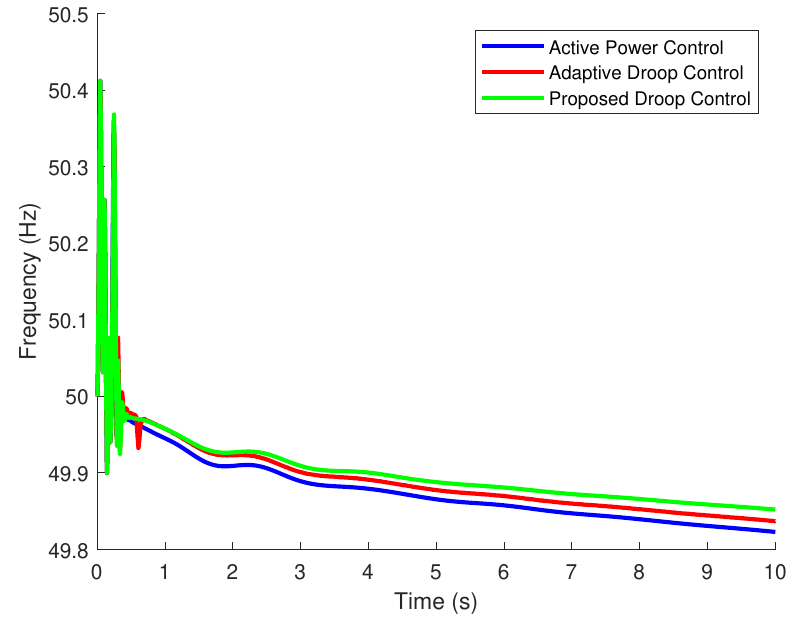}}
    \caption{AC frequencies in Scenario 1 under various control strategies.}
    \label{frequency_scenario1}
\end{figure}

\begin{figure}[!htbp]
    \centering
    \subfigure[DC voltages at MMC 1]{
        \includegraphics[width=0.48\textwidth]{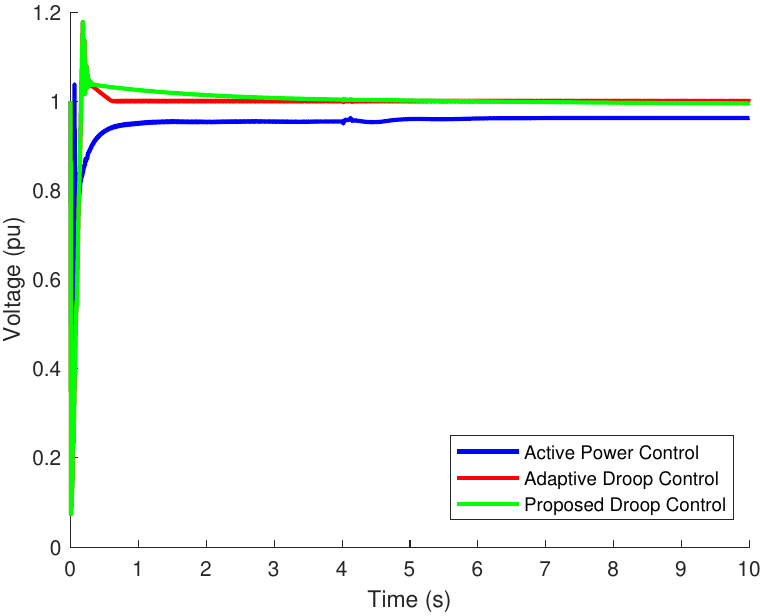}}%
    \hfill
    \subfigure[DC voltages at MMC 2]{
        \includegraphics[width=0.48\textwidth]{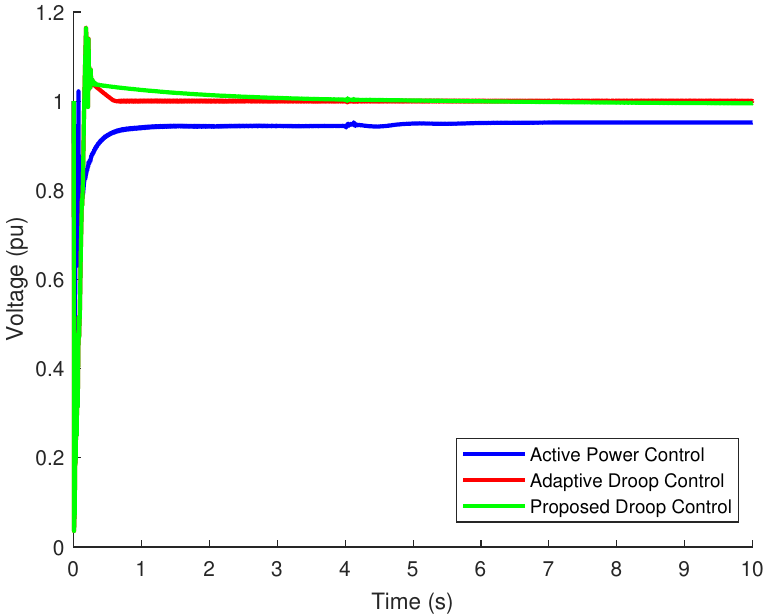}}
     \subfigure[DC voltages at MMC 3]{
        \includegraphics[width=0.48\textwidth]{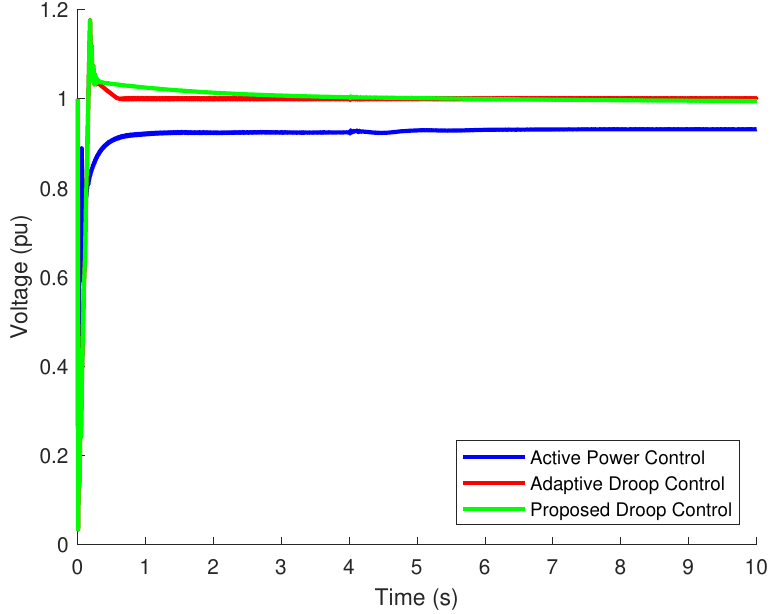}}
    \subfigure[DC voltages at MMC 4]{
        \includegraphics[width=0.48\textwidth]{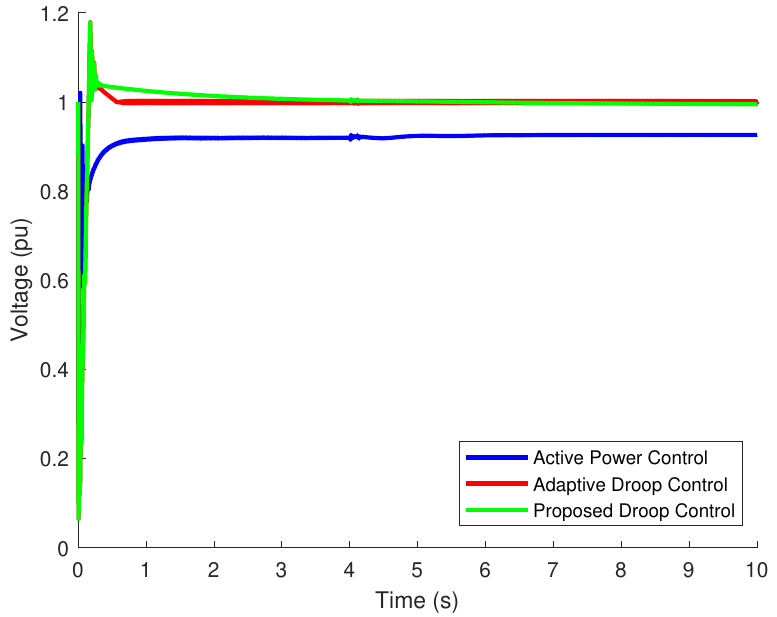}}
    \caption{DC voltages in Scenario 2 under various control strategies.}
    \label{voltage_scenario2}
\end{figure}

\begin{figure}[!htbp]
    \centering
    \subfigure[AC frequencies at MMC 1]{
        \includegraphics[width=0.48\textwidth]{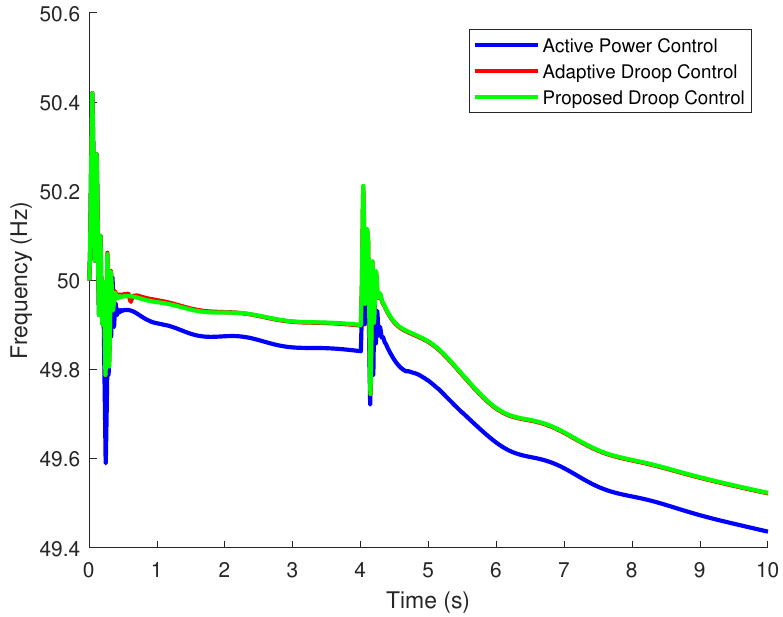}}
    \subfigure[AC frequencies at MMC 2]{
        \includegraphics[width=0.48\textwidth]{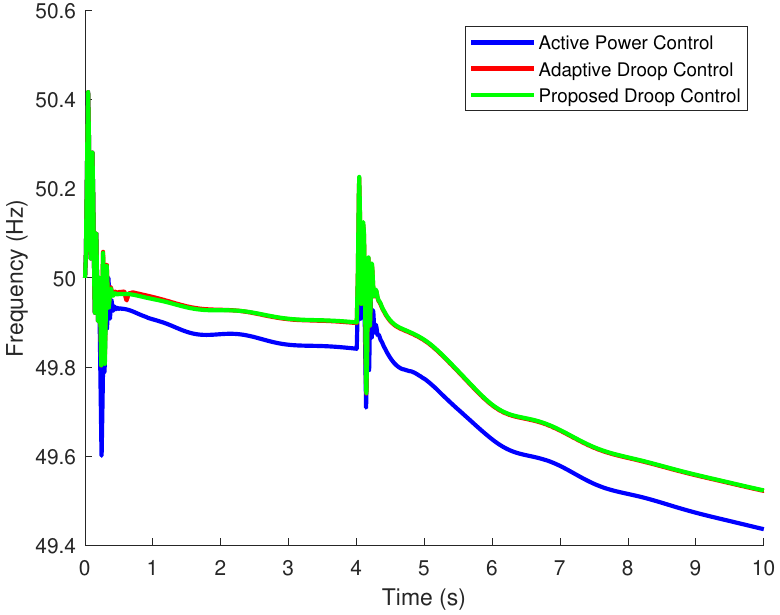}}
    \subfigure[AC frequencies at MMC 3]{
        \includegraphics[width=0.48\textwidth]{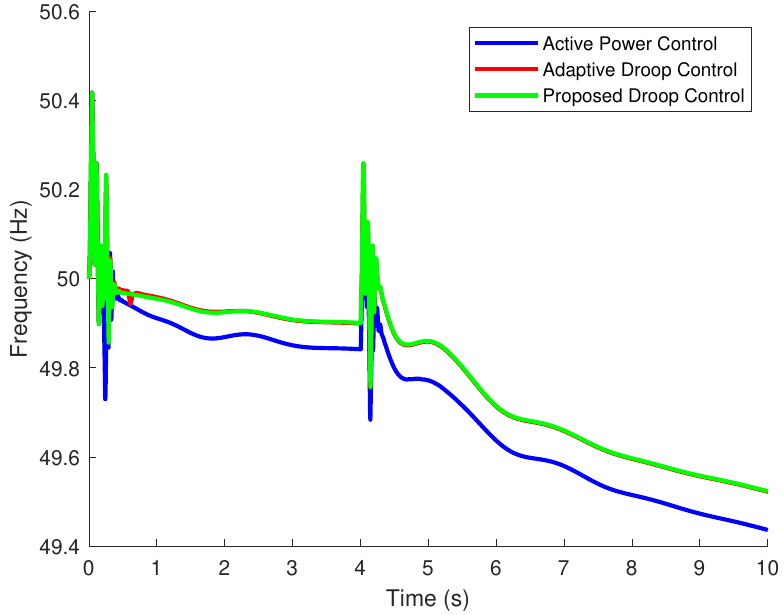}}
    \subfigure[AC frequencies at MMC 4]{
        \includegraphics[width=0.48\textwidth]{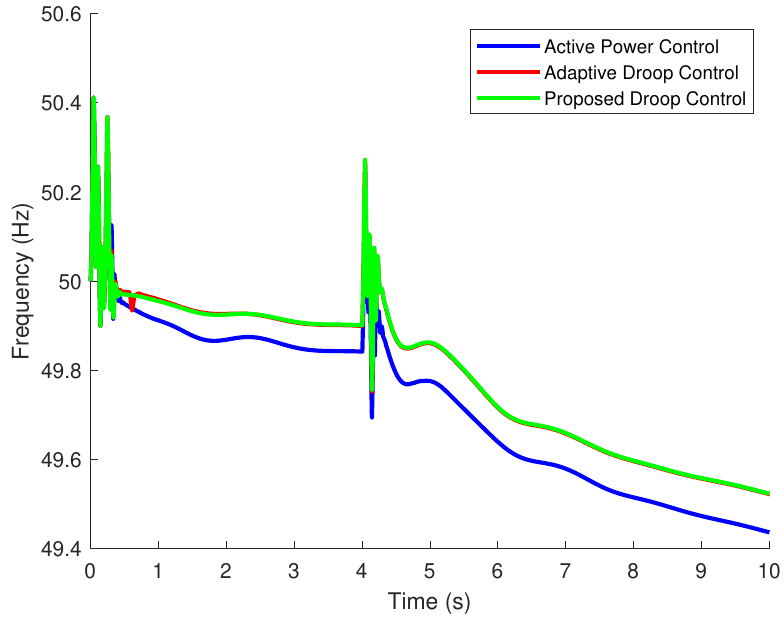}}
    \caption{AC frequencies in Scenario 2 under various control strategies.}
    \label{frequency_scenario2}
\end{figure}

\begin{figure}[!htbp]
    \centering
    \subfigure[DC voltages at MMC 1]{
        \includegraphics[width=0.48\textwidth]{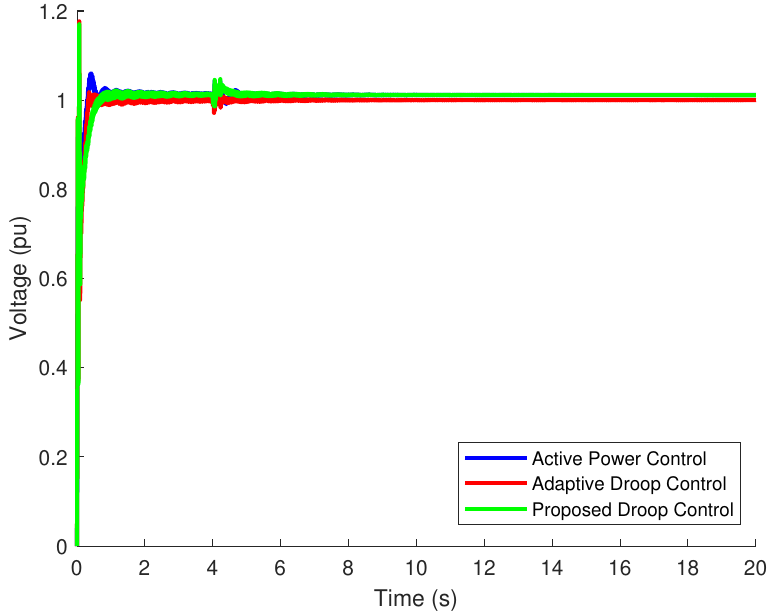}}%
    \hfill
    \subfigure[DC voltages at MMC 2]{
        \includegraphics[width=0.48\textwidth]{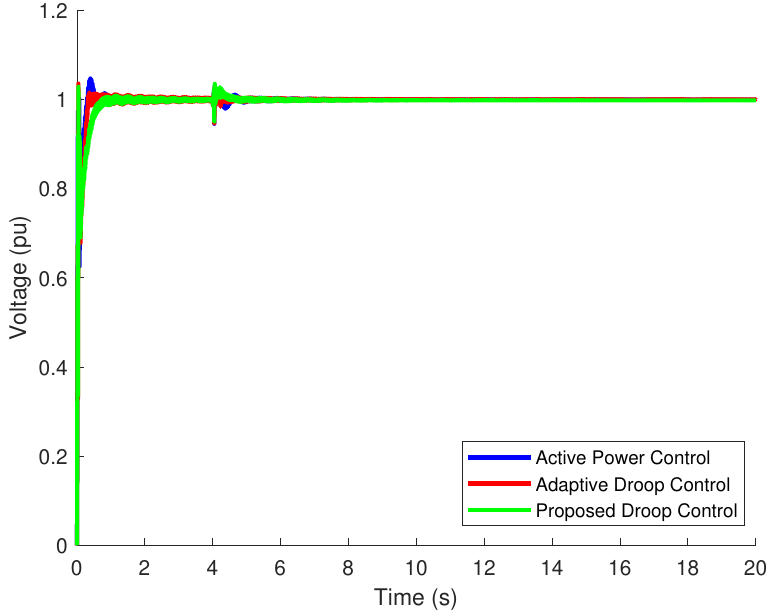}}
     \subfigure[DC voltages at MMC 3]{
        \includegraphics[width=0.48\textwidth]{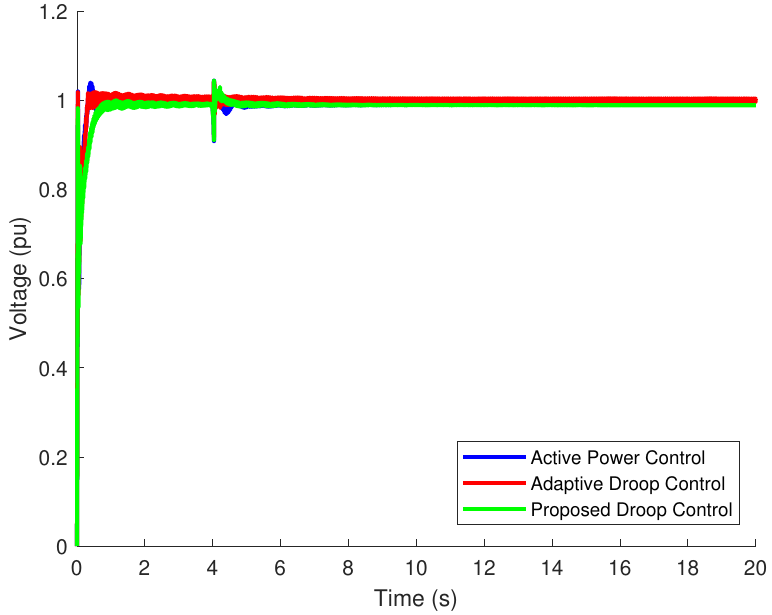}}
    \subfigure[DC voltages at MMC 4]{
        \includegraphics[width=0.48\textwidth]{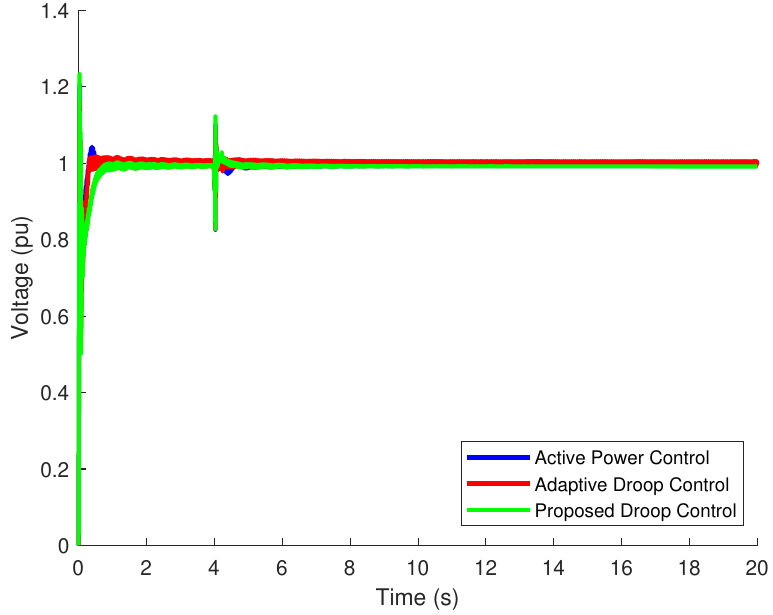}}
    \caption{DC voltages in Scenario 3 under various control strategies.}
    \label{voltage_scenario3}
\end{figure}

\begin{figure}[!htbp]
    \centering
    \subfigure[AC frequencies at MMC 1]{
        \includegraphics[width=0.48\textwidth]{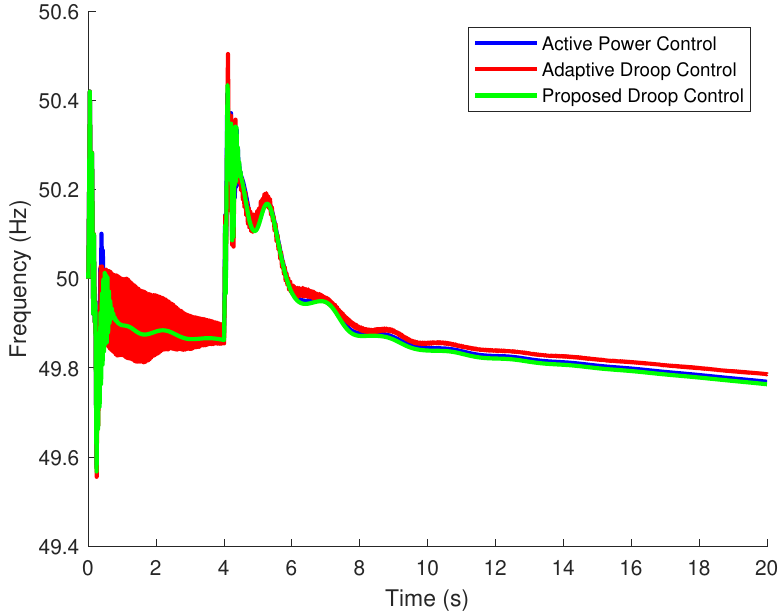}}
    \subfigure[AC frequencies at MMC 2]{
        \includegraphics[width=0.48\textwidth]{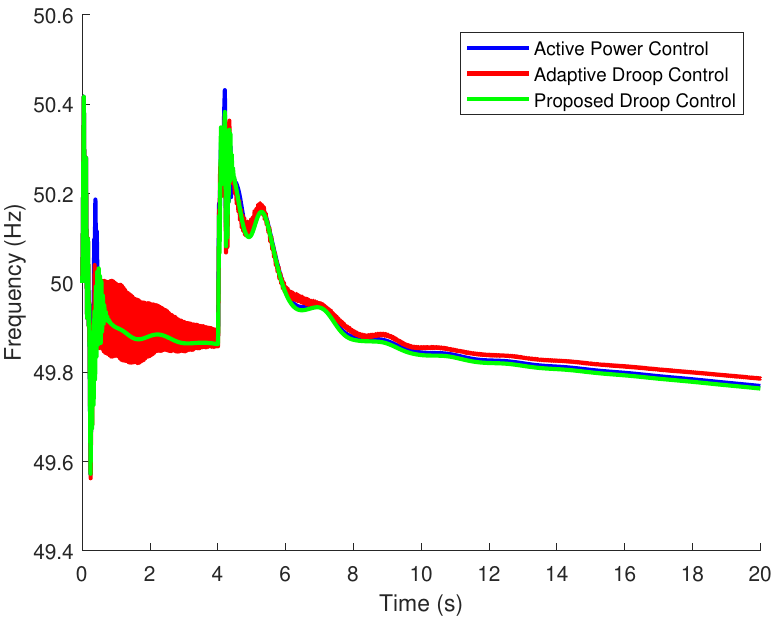}}
    \subfigure[AC frequencies at MMC 3]{
        \includegraphics[width=0.48\textwidth]{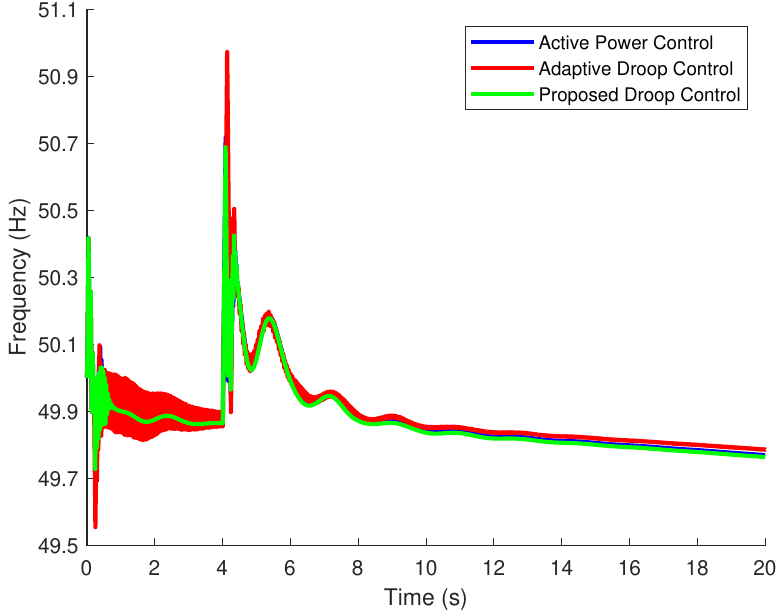}}
    \subfigure[AC frequencies at MMC 4]{
        \includegraphics[width=0.48\textwidth]{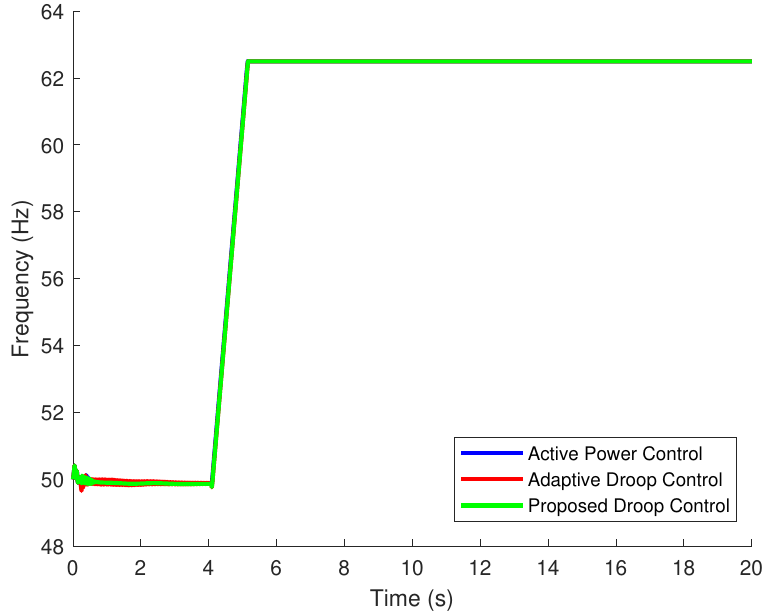}}
    \caption{AC frequencies in Scenario 3 under various control strategies.}
    \label{frequency_scenario3}
\end{figure}

To validate the effectiveness of the proposed droop control strategy, three predefined scenarios are simulated in EMTP. In Scenarios 2 and 3, system contingencies are introduced at $t_0 = 4.0s$, with the corresponding faults cleared after a duration of 0.2 seconds. The OPF results are used to initialize the operating points of the test system, and the converter control modes are configured accordingly. The simulation results, as illustrated in Figures \ref{voltage_scenario1}--\ref{frequency_scenario3} confirm that the proposed control scheme is capable of effectively mitigating the impact of faults in the AC/DC hybrid system. However, it should be noted that restoring full system stability solely through VSC control remains limited, particularly under severe contingencies. Achieving comprehensive stability may require the coordination of additional compensation components, which is beyond the scope of this study.

In Scenario 1, the system operates under stable conditions without any disturbances. As shown in Figure~\ref{voltage_scenario1}, both the adaptive droop control and the proposed method deliver comparable performance in voltage regulation. However, the active power control is insufficient to maintain voltage stability, leading to a continuous voltage rise. In terms of frequency regulation, as shown in Figure~\ref{frequency_scenario1}, the proposed method outperforms the alternatives by effectively damping oscillations and maintaining frequency values closer to the nominal setpoints at the PCC points.

In Scenario 2, the disconnection of Generator 16 results in a loss of 630 MW of power capacity. As illustrated in Figure~\ref{voltage_scenario2}, active power control fails to maintain voltage regulation, with voltages at all MMC terminals dropping to approximately 0.9 p.u. In contrast, the proposed control strategy achieves a more gradual voltage restoration process, restoring the voltages to their nominal values with minimal oscillations. This improved performance can be attributed to the dynamic adaptation of the droop parameters, which enhances the responsiveness of the system under stress. The frequency performance further underscores the limitations of conventional control. As shown in Figure~\ref{frequency_scenario2}, the active power control leads to larger and more prolonged frequency deviations. Meanwhile, the proposed method reduces both the magnitude and duration of frequency excursions, contributing to faster frequency stabilization. 

In Scenario 3, the disconnection of MMC 4 from the AC grid introduces a more severe contingency. To facilitate a thorough investigation, the simulation time is extended to 20 seconds. Under active power control, the system faces difficulties in converging after the contingency, suggesting that the inherent response characteristics of the active power control are insufficient for managing such a large disturbance. In response, MMC 2 is switched to constant voltage mode to provide voltage support, demonstrating performance comparable to the proposed droop control strategy. This highlights the strength of droop control in providing continuous support without requiring drastic adjustments to converter settings. Furthermore, as illustrated in Figure~\ref{frequency_scenario3}, the proposed control strategy enables faster frequency recovery with reduced oscillations following a disturbance. For instance, under adaptive droop control, the system frequency exhibits sustained oscillations and does not settle until 10 seconds after the disturbance. The oscillations are also more pronounced compared to the other schemes, indicating the limited damping capability of adaptive droop control under transient conditions. 

In summary, the OPF and time-domain simulation results across all three scenarios validate the effectiveness of the proposed control strategy in enhancing the stability and operational performance of the hybrid AC/DC system under various conditions. Compared with conventional active power control and adaptive droop control, the proposed method offers a more balanced, cost-efficient, and robust solution, particularly in maintaining voltage and frequency regulation during dynamic disturbances, which is a key factor for its practical deployment in real-world power systems. Nonetheless, the results also indicate that VSC-based control alone is insufficient to ensure full system stabilization under extreme contingencies. Therefore, the integration of supplementary compensation mechanisms is necessary to achieve comprehensive and resilient system stability.


\clearpage
\section{Conclusion}
This paper presents an optimal droop control strategy for hybrid AC-MTDC systems that enhances both voltage and frequency regulation, thereby improving power-sharing performance. The proposed strategy embeds a hierarchical OPF--based control framework and demonstrates greater adaptability and effectiveness in maintaining system stability across various scenarios, which is validated using a modified Nordic test system integrated with a four-terminal MTDC grid. Unlike conventional active power control or adaptive droop control, this approach integrates frequency control loops within MMCs, offering a more cost-efficient and robust solution for maintaining voltage and frequency during dynamic conditions and operational disturbances. Furthermore, by keeping voltage and frequency levels close to their nominal values, the strategy enables converters to maintain larger power margins, which is essential for managing significant power imbalances during contingencies without compromising system stability or reliability.

Future work will focus on refining the method to better accommodate uncertainties introduced by renewable energy sources, such as offshore wind. Incorporating these uncertainties into the optimization framework can further improve the control strategy’s resilience and performance under variable generation and demand conditions.

\clearpage
\bibliographystyle{elsarticle-num}
\bibliography{manual.bib} 

\end{document}